\documentclass[pra,twocolumn,groupedaddress,showpacs,floatfix]{revtex4}

\usepackage{graphics}
\usepackage{graphicx}
\usepackage{bm}
\usepackage{amsmath}
\usepackage{amsfonts}
\usepackage{amssymb}
\usepackage{latexsym}
\usepackage{color}

\begin{document}

\title{Comparing cavity and ordinary laser cooling within the Lamb-Dicke regime}
\author{Tony Blake,\footnote{Corresponding author: pytb@leeds.ac.uk} Andreas Kurcz, and Almut Beige}
\affiliation{The School of Physics and Astronomy, University of Leeds, Leeds, LS2 9JT, United Kingdom} 

\date{\today}

\begin{abstract}
Cavity-mediated cooling has the potential to become one of the most efficient techniques to cool molecular species down to very low temperatures. In this paper we analyse cavity cooling with single-laser driving for relatively large cavity decay rates $\kappa$ and relatively large phonon frequencies $\nu$. It is shown that cavity cooling and ordinary laser cooling are essentially the same within the validity range of the Lamb-Dicke approximation. This is done by deriving a closed set of rate equations and calculating the corresponding stationary state phonon number and cooling rate. For example, when $\nu$ is either much larger or much smaller than $\kappa $, the minimum stationary state phonon number scales as $\kappa^2/16 \nu^2$ (strong confinement regime) and as $\kappa / 4 \nu$ (weak confinement regime), respectively.
\end{abstract}


\maketitle

\section{Introduction} \label{Intro}

One motivation for cooling particles by coupling them to an optical cavity is the hope to cool many of them simultaneously and very efficiently to very low temperatures \cite{domokos2,cool}. Beyond this, cavity cooling might open the possibility to transfer not only atoms and ions but also molecules to very low phonon numbers \cite{mol,andre,andre2,andre3,mol2}. In general, molecules do not possess the closed transitions \cite{DeMille} required for laser cooling \cite{sideband,Stenholm2,sideband2,ions}. Using for example ordinary laser cooling, the spontaneous emission of photons would rapidly populate states where the particles no longer see the cooling laser. Cavity cooling overcomes this problem by depopulating the excited electronic states of molecules in a two-step process \cite{peter2}. First, the population in the excited state is transferred into the cavity mode. From there it leaks into the environment via spontaneous emission.

Indications that cavity couplings might enhance the confinement of trapped particles have been observed as early as 1997 in Grangier's group in Paris \cite{Grangier}. Systematic experimental studies of cavity cooling have subsequently been reported by the group of Rempe \cite{rempe00,pinkse2,Rempe,rempe09}, Vuleti\'c \cite{vuletic23,vuletic,vul}, and others \cite{kimble,chap}. Recent atom-cavity experiments access an even wider range of experimental parameters by using optical ring cavities \cite{Nagorny,Nagorny2} and by combining optical cavities with atom chip technology \cite{trupke,reichel}, atomic conveyer belts \cite{meschede0,meschede}, and ion traps \cite{Drewsen}. However, these experiments are still limited to the cooling of atoms and ions. Moreover, when it comes to the cooling of large numbers of particles, evaporative and sympathetic cooling still seems a much more efficient approach \cite{evaporative,evaporative2,sympathetic,sympathetic2}.

\begin{figure}[t]
\begin{minipage}{\columnwidth}
\begin{center}
\includegraphics[scale=0.6]{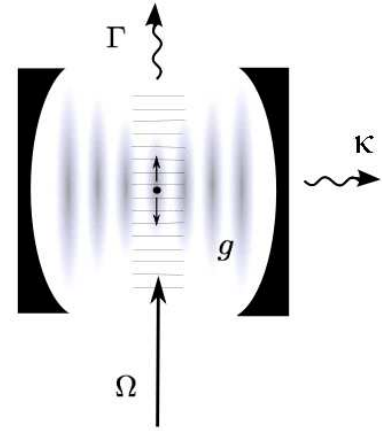}
\end{center}
\caption{Experimental setup of a single two-level particle inside an optical cavity with coupling constant $g$ and spontaneous decay rates $\kappa$ and $\Gamma$. The motion of the particle orthogonal to the cavity axis is strongly confined by an externally applied harmonic trapping potential with phonon frequency $\nu$. The cooling of this vibrational mode can be done with the help of the cooling laser with Rabi frequency $\Omega$.} \label{setup}
\end{minipage}
\end{figure}

Theoretical work on cavity cooling of strongly confined particles has been pioneered by Zoller's group in 1993. In Refs.~\cite{Cirac2,Cirac4}, they calculate the cooling rate and the final temperature of a single two-level particle trapped inside an optical resonator with the help of a quantum optical master equation within the Lamb-Dicke regime. A similar master equation approach to cavity cooling has been used in Refs.~\cite{cool,morigi,morigi2}. Cavity cooling of free particles was first discussed in Refs.~\cite{lewen,lewen2}. Later, Ritsch and collaborators \cite{domokos2,peter2,ritsch97,ritsch98,peter3}, Vuleti{\'c} {\em et al.} \cite{vuletic10,vuletic3}, Murr {\em et al.} \cite{Murr,Murr2,Murr3}, and others \cite{Tan} developed semiclassical theories to model cavity cooling processes, including the cooling of polarisable particles \cite{nano,nano2,nano3}. Moreover, Xuereb {\em et al.} \cite{Xuereb} introduced a simple input-output formalism which can in principle be applied to a variety of cooling scenarios. 

This paper summarises and generalises our current knowledge about cavity cooling within the so-called Lamb-Dicke regime \cite{cool,Cirac2,Cirac4,morigi,morigi2,vuletic3} and compares cavity cooling with ordinary laser cooling. Moreover, it introduces a methodology which paves the way to an analysis of cavity cooling beyond the Lamb-Dicke approximation \cite{tony2}. The setup considered in this paper is shown in Figure \ref{setup} and consists of a single two-level particle with ground state $|0 \rangle$ and excited state $|1 \rangle$ trapped inside an optical cavity. In this paper, we assume confinement of the motion of the particle in the direction of the cooling laser which enters the setup orthogonal to the cavity axis. The aim of the described cooling process is to minimise the mean phonon number in this vibrational mode.

\begin{figure}
\begin{center}
\includegraphics[scale=0.2]{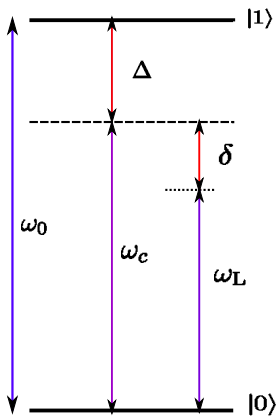}
\end{center}
\caption{Level configuration showing a single particle with ground $|0 \rangle$ and excited state $|1 \rangle$. Here $\omega_{\rm c}$ and $\omega_{\rm L}$ are the frequency of the cavity and of the cooling laser. The corresponding detunings with respect to frequency of the 0--1 transition, $\omega_0$, are $\Delta$ and $\Delta + \delta$.} \label{energy1}
\end{figure}

The energy levels considered in this paper are shown in Figure \ref{energy1}. In the following, we denote the detuning of the cavity and of the laser with respect to the 0--1 transition of the particle by $\Delta$ and $\Delta + \delta$, respectively. Here we are especially interested in the case, where $\Delta $ is much larger than all other system parameters. In this case, the particle remains predominantly in its ground state and its electronic states can be adiabatically eliminated from the time evolution of the system. As we shall see below, spontaneous emission from the excited atomic state remains negligible throughout the cooling process, when the experiment operates well within the so-called strong coupling regime, i.e.~when the cavity coupling constant $g^2$ is much larger than the product of the spontaneous decay rates $\kappa$ and $\Gamma$. Outside, the strong coupling regime, the cooling process illustrated in Figure \ref{setup} becomes a mixture of cavity and ordinary laser cooling \cite{vuletic3}. 

In the following, we use the quantum optical master equation for atom-cavity systems \cite{Cirac2,Cirac4} to derive a closed set of rate equations. These equations are linear differential equations which describe the time evolution of expectation values. Only two of the variables in these cooling equations are populations: the mean phonon number $m$ and the mean number of photons in the cavity $n$. All other variables are coherences. More concretely, we assume that the mean phonon number $m$ evolves on a much slower time scale than all the other expectation values which are included in the cooling equations, as it applies for a very wide range of experimental parameters. This allows us to simply reduce the above mentioned cooling equations to a single effective cooling equation via an adiabatic elimination of all expectation values other than $m$. As a result, we obtain the cooling rate and the stationary state phonon number as a function of the system parameters. 

Different from Refs.~\cite{Cirac4,morigi,morigi2} we do not restrict ourselves to the case where the detuning $\delta$ is close to the phonon frequency $\nu$. Here we give analytical results for the stationary state phonon number $m^{\rm ss}$ of the trapped particle for the general case. Different from Refs.~\cite{vuletic,domokos2,peter2,ritsch97,ritsch98,peter3,vuletic10,vuletic3,Murr,Murr2,Murr3,Tan,nano,nano2,nano3}, we avoid semiclassical approximations. Different from Ref.~\cite{cool}, we avoid the rotating wave approximation with respect to the phonon frequency $\nu$. This allows us to calculate cooling rates and stationary state phonon numbers correctly even when the phonon frequency $\nu$ is relatively small. In fact, it can be shown that the counter-rotating terms in the particle-phonon interaction can heat a system very rapidly \cite{SL}.

The general idea behind cavity cooling is the continuous conversion of phonons into cavity photons. While phonons have no decay rate, photons leak out through the cavity mirrors easily. The conversion of phonons into photons can hence result in the constant removal of phonon energy from the system. The role of the cooling laser is to establish an effective coupling between the vibrational states of the particle and the cavity field. Cooling, i.e.~the permanent loss of a phonon, occurs when the absorption of a phonon is accompanied by the creation of a photon in the cavity field which subsequently leaks into the environment. 

As we shall see below, in the validity range of the so-called Lamb-Dicke approximation, there are many similarities between cavity and ordinary laser cooling. For example, it will be shown that
one should choose the effective laser detuning 
\begin{eqnarray} \label{deff}
\delta_{\rm eff} &\equiv & \delta - {g^2 \over \Delta}
\end{eqnarray}
equal to $\nu$, when the cavity decay rate $\kappa$ is much smaller than the phonon frequency $\nu$ (strong confinement regime) in order to minimise the final number of phonons in the system. The corresponding stationary state phonon number $m^{\rm ss}_{\delta_{\rm eff} = \nu}$ will be found to be approximately given by $\kappa^2/16 \nu^2$. However, if $\kappa$ is much larger than $\nu$ (weak confinement regime), one should choose $\delta_{\rm eff} = {1 \over 2} \kappa$, since $m^{\rm ss}_{\delta_{\rm eff} = {1 \over 2} \kappa}$ equals $\kappa/4 \nu$ to a very good approximation. Comparing the explicit analytical expressions for $m^{\rm ss}_{\delta_{\rm eff} = {1 \over 2} \kappa} $ and $m^{\rm ss}_{\delta_{\rm eff} = \nu}$, we find that 
\begin{eqnarray} \label{last}
m^{\rm ss}_{\delta_{\rm eff} = {1 \over 2} \kappa} &=& \sqrt{m^{\rm ss}_{\delta_{\rm eff} = \nu}} 
\end{eqnarray}
for a wide range of experimental parameters. This means, when the trapped particle cannot be so strongly confined that it is {\em not} possible to cool to phonon numbers below one, it is better to choose $\delta_{\rm eff} = {1 \over 2} \kappa$ than choosing $\delta_{\rm eff} = \nu$, since $m^{\rm ss}_{\delta_{\rm eff} = {1 \over 2} \kappa} < m^{\rm ss}_{\delta_{\rm eff} = \nu}$ in this case.

Comparing the explicit analytical expressions for the cooling rates $\gamma_{\delta_{\rm eff} = \nu}$ and $\gamma_{\delta_{\rm eff} = {1 \over 2} \kappa}$ for the two above mentioned cases, i.e.~the strong and the weak confinement regime, we will find that
\begin{eqnarray} \label{gammadouble2}
{\gamma_{\delta_{\rm eff} = {1 \over 2} \kappa} \over \gamma_{\delta_{\rm eff} = \nu}} &=& {\kappa \over 8 \nu } 
\end{eqnarray}
to a very good approximation. This means, choosing $\delta_{\rm eff}= {1 \over 2} \kappa$ instead of $\delta_{\rm eff}= \nu$ yields a speed up of the cooling process when $\kappa > 8 \nu$. As we have seen in the previous paragraph, this detuning is also the better choice in order to minimise the stationary state phonon number in the weak confinement regime. Speeding up the cooling process is important, since spontaneous emission from the excited state $|1 \rangle$ remains thus negligible for a much wider range of single particle-cavity cooperativity parameters $g^2 / \kappa \Gamma$. 

There are five sections in this paper. In Section \ref{model} we derive the master equation for the experimental setup shown in Figure \ref{setup}. The main approximations made in this section are the Lamb-Dicke approximation,  the rotating wave approximation with respect to optical transitions, and the adiabatic elimination of the electronic states of the particle. Section \ref{standard} analyses the cavity cooling process. Starting from the master equation, we obtain a full set of cooling equations which are then used to calculate the stationary state phonon number $m^{\rm ss}$ and the cooling rate $\gamma$ as a function of the experimental parameters. Section \ref{new} compares cavity cooling with ordinary laser cooling. Finally, we summarise our results in Section \ref{conc}. There we emphasize that going beyond the Lamb-Dicke regime might yield very different results. 
 
\section{Theoretical model} \label{model}

Let us start by introducing the theoretical model for the description of the experimental setup in Figure \ref{setup}. It contains a strongly confined particle coupled to an optical cavity. The cooling laser drives the electronic transition of the particle and enters the setup in a direction orthogonal to the cavity axis. In the following, we assume that an external trapping potential confines the particle in the laser direction such that its motion becomes quantised. The aim of the cavity cooling scheme discussed in this paper is to minimise the number of phonons in this mode. Cooling other vibrational modes of the particle motion would require additional cooling lasers.

\subsection{The Hamiltonian}

The Hamiltonian of the system in Figure \ref{setup} can be written as 
\begin{eqnarray} \label{1.23} 
H &=& H_{\rm par} + H_{\rm phn} + H_{\rm cav} + H_{\rm L} + H_{\rm par-cav} \, .
\end{eqnarray}
The first three terms are the energy of the electronic states of the trapped particle, its quantised vibrational mode, and the quantised cavity field mode. Suppose, the particle is effectively a two-level system with ground state $|0\rangle$ and excited state $|1\rangle$ and the energies $\hbar \omega_0$, $\hbar \nu$, and $\hbar \omega_{\rm c}$ are the energy of a single atomic excitation, a single phonon, and a single cavity photon, respectively. Then
\begin{eqnarray} \label{1.23b}
H_{\rm par} &=& \hbar \omega_0 \, \sigma^+ \sigma^- \, , \notag \\
H_{\rm phn} &=& \hbar \nu \, b^\dagger b \, , \notag \\
H_{\rm cav} &=& \hbar \omega_{\rm c} \, c^\dagger c \, ,
\end{eqnarray}
where the operators $\sigma^{-} \equiv |0\rangle\langle1|$ and $\sigma^{+}\equiv|1\rangle\langle0|$ are the atomic lowering and raising operators, and $b$ and $c$ are the phonon and the photon annihilation operators with the commutator relations
\begin{eqnarray} \label{comms}
[b,b^{\dagger}]=[c,c^{\dagger}]=1 \, . 
\end{eqnarray}
The two remaining terms in Eq.~(\ref{1.23}), i.e.~$H_{\rm L}$  and $H_{\rm par-cav}$, are the laser Hamiltonian and the Hamiltonian describing the interaction between the trapped particle and the cavity mode. Let us now have a closer look at these terms.
 
The role of the cooling laser is to establish a coupling between the electronic states $|0 \rangle$ and $|1 \rangle$ of the trapped particle and its quantised motion. In the dipole approximation, its Hamiltonian can be written as
\begin{eqnarray} \label{HL1}
H_{\rm L} &=& e\textbf{D} \cdot \textbf{E}_{\rm L} ({\bf x},t) \, ,
\end{eqnarray}
where $e$ is the charge of an electron, ${\bf D}$ is the atomic dipole moment, and $ \textbf{E}_{\rm L} (\textbf{x},t)$ is the electric field of the laser at the position ${\bf x}$ of the particle relative to its equilibrium position ${\bf R} = 0$ at time $t$. More concretely, the dipole moment ${\bf D}$ can be written as
\begin{eqnarray} \label{DDD}
\textbf{D} &=& \textbf{D}_{01} \, \sigma^{-}+\mbox{H.c.} \, ,
\end{eqnarray}
where $\textbf{D}_{01}$ is a 3-dimensional complex vector, and  
\begin{eqnarray}
\textbf{E}_{\rm L} (\textbf{x},t)=\textbf{E}_0 \, {\rm e}^{{\rm i}(\textbf{k}_{\rm L} \cdot\textbf{x}-\omega_{\rm L} t)} + {\rm c.c.} 
\end{eqnarray} 
with $\textbf{E}_0$, $\textbf{k}_{\rm L}$, and $\omega_{\rm L}$ being the amplitude, the wave vector, and the frequency of cooling laser.  

As already mentioned above, the cooled motion of the trapped particle is its center of mass motion in the laser direction. Considering this motion as quantised with the phonon annihilation operator $b$ from above, yields
\begin{eqnarray} \label{2.21}
\textbf{k}_{\rm L} \cdot \textbf{x} &=& \eta (b + b^\dagger) \, , 
\end{eqnarray}
where the Lamb-Dicke parameter $\eta $ is a measure for the steepness of the trapping potential \cite{Stenholm2}. Introducing the particle displacement operator $D$ as 
\begin{eqnarray}\label{2.22}
D ({\rm i}\eta) &\equiv & {\rm e}^{- {\rm i} \eta (b+b^\dagger)} \, ,
\end{eqnarray}
and substituting Eqs.~(\ref{DDD})--(\ref{2.22}) into Eq.~(\ref{HL1}), we see that the laser Hamiltonian can also be written as
\begin{eqnarray}\label{2.4}
H_{\rm L} &=& e \left[ \textbf{D}_{01} \, \sigma^{-} + {\rm H.c.} \right] \cdot \textbf{E}_0^* \, D({\rm i}\eta) \, {\rm e}^{ {\rm i} \omega_{\rm L} t} + {\rm H.c.} ~~~~
\end{eqnarray}
The cooling laser indeed couples vibrational and electronic states of the particle.

The interaction Hamiltonian describing the coupling between the electronic states of the trapped particle and the cavity equals in the usual dipole approximation  
\begin{eqnarray}
H_{\rm par-cav} &=& e\textbf{D}\cdot\textbf{E}_{\rm cav}(\textbf{x}) \, ,
\end{eqnarray}
where $\textbf{E}_{\rm cav}(\textbf{x})$ is the observable for the quantised electric field inside the resonator. Denoting the coupling constant between the particle and the cavity field as $g$, this Hamiltonian can be written as
\begin{eqnarray}\label{2.10}
H_{\rm par-cav} &=& \hbar g (\sigma^{-}+\sigma^{+}) \, c+ {\rm H.c.} 
\end{eqnarray} 
which describes the possible exchange of energy between atomic states and the cavity mode with a constant cavity coupling constant $g$.

\subsection{Interaction picture}

We now change into an interaction picture which allows us to take advantage of the fact that the experimental parameters $\Omega$, $\delta$, $\nu$, $g$, and $\Delta $ are in general much smaller than the optical frequencies, i.e. 
\begin{eqnarray} \label{Delta}
\Omega \, , \, \delta \, , \, \nu \, , \, g \, , \, \Delta  &\ll & \omega_{\rm L} \, , ~ \omega_{\rm c} \, . 
\end{eqnarray}
Choosing 
\begin{eqnarray} \label{intpic}
H_0 &=& \hbar\omega_{\rm L} \, \sigma^{+}\sigma^{-}+\hbar \left( \omega_{\rm c} - \delta \right) \, c^{\dagger}c \, ,
\end{eqnarray}
we find that the interaction Hamiltonian $H_{\rm I}$,
\begin{eqnarray}
H_{\rm I}=U^{\dagger}_0(t,0) \, (H-H_0) \, U_0(t,0) \, ,
\end{eqnarray}
contains terms which oscillate with frequencies close to $2 \omega_{\rm L}$, $2 \omega_{\rm c}$, $\omega_{\rm L} + \omega_{\rm c}$. Neglecting these relatively fast oscillating terms as part of the usual rotating wave approximation, $H_{\rm I}$ becomes 
\begin{eqnarray}\label{2.8}
H_{\rm I} &=& {1 \over 2} \hbar \Omega \, D({\rm i}\eta) \sigma^- + \hbar g \, \sigma^- c^+ +\mbox{H.c.}  \notag \\
&& + \hbar\left(\Delta+\delta \right)\sigma^+\sigma^- +\hbar\nu \, b^{\dagger}b +\hbar\delta \, c^{\dagger}c \, , 
\end{eqnarray}
where 
\begin{eqnarray}
\Omega &\equiv & {2e \over \hbar} \, \textbf{D}_{01} \cdot \textbf{E}_0^* 
\end{eqnarray}
denotes as usual the laser Rabi frequency. This Hamiltonian is time-independent and will be used later to analyse the time evolution of the mean phonon number. 

\subsection{Adiabatic elimination of the electronic states} \label{adel}

To minimise spontaneous emission from the excited electronic state of the trapped particle, i.e.~to maintain a closed cooling cycle, we assume in the following that the detuning $\Delta$ is much larger than all other system parameters,  
\begin{eqnarray} \label{condi}
\Delta &\gg& \Omega \, , ~ \delta \, ,~ \nu \, , ~ g \, , ~ \Gamma \, , ~ \kappa \, .
\end{eqnarray}
Concentrating on this parameter regime allows us to eliminate the particle adiabatically from the time evolution. To do so, we write the state vector of the system as
\begin{eqnarray}
|\psi\rangle= \sum_{j=0}^1 \sum_{m,n=0}^\infty c_{jmn} \, |jmn\rangle \, ,
\end{eqnarray} 
where $|j \rangle$ denotes the atomic state and $|m \rangle$ and $|n \rangle$ are phonon and photon number states, respectively. According to the Schr\"odinger equation, the time evolution of the coefficient $c_{j'm'n'}$ is given by
\begin{eqnarray}\label{3.3}
\dot{c}_{j'm'n'} &=& - {{\rm i} \over \hbar} \sum_{j=0}^1 \sum_{m,n=0}^\infty c_{jmn}\langle j'm'n'| H_{\rm I} |jmn\rangle \, . ~~
\end{eqnarray}
Given condition (\ref{condi}), the $c_{j'm'n'}$ with $j'=1$ evolve on a much faster time scale than the coefficients with $j'=0$. Setting their derivatives equal to zero, we find
\begin{eqnarray} \label{c1-c0} 
c_{1m' n'} &=& - \frac{1}{2 \Delta} \sum_{m,n=0}^\infty c_{0mn} \notag \\ 
&& \times \langle m' n' | \, \left[ 2g \, c + \Omega \, D^{\dag}({\rm i}\eta) \right] \, |mn \rangle  ~~~
\end{eqnarray}
which holds up to first order in $1/\Delta$. Substituting this result into the differential equations for $c_{0m'n'}$, we obtain the effective interaction Hamiltonian
\begin{eqnarray} \label{3.7}
H_{\rm I} &=& \hbar g_{\rm eff} \, D({\rm i}\eta) c + {\rm H.c.} + \hbar\nu \, b^{\dagger}b + \hbar \delta_{\rm eff} \, c^{\dagger} c ~~~
\end{eqnarray}
with the (real) cavity coupling constant $g_{\rm eff}$ given by 
\begin{eqnarray} \label{3.8} 
g_{\rm eff} &\equiv & - \frac{g \Omega}{2\Delta} 
\end{eqnarray}
and the effective detuning $\delta_{\rm eff}$ as defined in Eq.~(\ref{deff}). The interaction Hamiltonian $H_{\rm I}$ in Eq.~(\ref{3.7}) holds up to first order in $1/\Delta$. It no longer contains any atomic operators. Instead, Eq.~(\ref{3.7}) describes a direct interplay between phonons and cavity photons. 

\subsection{Lamb-Dicke approximation} \label{simple}

Suppose the particle has already been cooled enough to ensure that it remains in the vicinity of its equilibrium position ${\bf R} = 0$. More concretely, we assume in the following that the displacement ${\bf x}$ of the particle is small compared to the wavelength of the cooling laser. Then $\textbf{k}_{\rm L} \cdot \textbf{x}$ in Eq.~(\ref{2.21}) is much smaller than one, and the so-called Lamb-Dicke approximation with 
\begin{eqnarray} \label{condi0}
\eta & \ll & 1 
\end{eqnarray}
can be applied. This means, Eq.~(\ref{2.22}) simplifies to
\begin{eqnarray} \label{2.22b}
D ({\rm i}\eta) &=& 1- {\rm i}\eta(b+b^{\dagger}) \, .
\end{eqnarray} 
Substituting this into Eq.~(\ref{3.7}), we finally obtain the interaction Hamiltonian
\begin{eqnarray} \label{3.7b}
H_{\rm I} &=& \hbar g_{\rm eff} \, c - {\rm i} \hbar \, \eta g_{\rm eff} \, (b + b^\dagger) c + \mbox{H.c.} + \hbar \nu \, b^{\dagger} b  \notag \\
&& + \hbar \delta_{\rm eff} \, c^{\dagger}c 
\end{eqnarray}
which contains cavity interactions, the phonon-photon interaction, the phonon energy term, and a level shift.

\subsection{Master equation}

After the adiabatic elimination of the electronic states of the particle, the only relevant decay channel in the system is the leakage of photons through the cavity mirrors. To take this into account, we describe the cooling process in the following by the master equation in the usual Lindblad form
\begin{eqnarray}\label{3.1}
\dot{\rho}= - {{\rm i} \over \hbar} \left[H_{\rm I},\rho\right] + {1 \over 2} \kappa \left( 2 \, c \rho c^\dagger - c^\dagger c \rho -\rho c^\dagger c \right) 
\end{eqnarray}
with $H_{\rm I}$ as in Eq.~(\ref{3.7b}). 

\section{Analysis of the cooling process} \label{standard}

In this section, we calculate the stationary state phonon number $m^{\rm ss}$ and the effective cooling rate $\gamma$ as a function of the experimental parameters $\eta$, $g_{\rm eff}$, $\kappa$, $\nu$, and $\delta_{\rm eff}$. This is done by using the above master equation to obtain a closed set of cooling equations which predict the time evolution of the mean phonon number $m$. In the following, we are especially interested in the case where the cavity decay rate $\kappa$ and the phonon frequency $\nu$ are both relatively large. More concretely, we assume in the following that  
\begin{eqnarray} \label{condiA}
\kappa , \, \nu &\gg & \eta g_{\rm eff} \, .
\end{eqnarray}
As we shall see below, the mean phonon number $m$ evolves in this case on a much slower time scale than all other expectation values which are included in the cooling equations. The latter can hence be eliminated adiabatically from the time evolution of the system.

\subsection{Cooling equations} \label{cooling2}

Since the time derivative of the expectation value of an operator $A$ equals 
\begin{eqnarray}
\dot {\langle A \rangle} &=& \mbox{Tr} (A\dot{\rho}) \, ,
\end{eqnarray}
the master equations in Eq.~(\ref{3.1}) implies
\begin{eqnarray} \label{dotA}
\langle \dot A \rangle &=& -{{\rm i} \over \hbar} \, \left\langle\left[A,H_{\rm I}\right]\right\rangle + {1 \over 2} \kappa \, \langle 2 \,  c^\dagger A c - A c^\dagger c - c^\dagger c A \rangle \, . ~~~~~
\end{eqnarray}
Using the commutator relations in Eq.~(\ref{comms}) and the interaction Hamiltonian in Eq.~(\ref{3.7b}) and applying this equation to the mean phonon number $m = \langle b^{\dagger}b\rangle$ and the mean photon number $n = \langle c^{\dagger}c \rangle$, and the coherences
\begin{eqnarray} \label{coherences}
&& k_{\rm x} = {\rm i} \langle b-b^{\dagger} \rangle \, , ~~
k_{\rm y} ={\rm i} \langle c-c^{\dagger} \rangle \, , \notag \\
&& k_{\rm u} =\langle b+b^{\dag}\rangle \, , ~~ 
k_{\rm w} =\langle c+c^{\dag}\rangle \, , \notag\\
&& k_1 =\langle(b+b^{\dag})(c+c^{\dag})\rangle \, , ~~
k_2 ={\rm i}\langle(b+b^{\dag})(c-c^{\dag}) \rangle \, , \notag\\  
&& k_3 ={\rm i}\langle(b-b^{\dag})(c+c^{\dag})\rangle \, , ~~
k_4 = \langle (b - b^{\dag})(c - c^\dagger) \rangle \, , \notag \\
&& k_5 = \langle c^2+c^{\dag 2}\rangle \, , ~~ 
k_6 ={\rm i}\langle c^2- c^{\dag 2}\rangle \, , \notag \\
&& k_7 =\langle b^2+b^{\dag 2}\rangle \, , ~~ 
k_8 ={\rm i}\langle b^2-b^{\dag 2}\rangle 
\end{eqnarray}
we obtain a closed set of differential equations. These are 
\begin{eqnarray} \label{48}
\dot k_{\rm x} &=& - 2 \eta g_{\rm eff} \, k_{\rm y} + \nu \, k_{\rm u}  \, , \notag \\
\dot k_{\rm y} &=& 2  g_{\rm eff} + \delta_{\rm eff} \, k_{\rm w} - {1 \over 2} \kappa \, k_{\rm y} \, , ~~~\notag \\
\dot k_{\rm u} &=& - \nu \, k_{\rm x} \, , \notag \\
\dot k_{\rm w} &=& 2 \eta g_{\rm eff} \, k_{\rm u} - \delta_{\rm eff} \, k_{\rm y} - {1 \over 2} \kappa \, k_{\rm w}
\end{eqnarray}
and 
\begin{eqnarray} \label{49}
\dot n &=& g_{\rm eff} \, k_{\rm y} + \eta g_{\rm eff} \, k_{\rm 1} - \kappa \, n \, , \notag \\
\dot k_1 &=& 2 \eta g_{\rm eff} \left( k_7 + 2m + 1 \right) - \nu \, k_3 - \delta_{\rm eff} \, k_2 - {1 \over 2} \kappa \, k_1 \, , \notag \\
\dot k_2 &=& 2  g_{\rm eff} \, k_{\rm u} + \nu \, k_4 + \delta_{\rm eff} \, k_1 - {1 \over 2} \kappa \, k_2 \, , \notag \\
\dot k_3 &=&  - 2 \eta g_{\rm eff} \, ( k_6 - k_8) + \nu \, k_1 + \delta_{\rm eff} \, k_4 - {1 \over 2} \kappa \, k_3 \, , \notag \\
\dot k_4 &=& - 2  g_{\rm eff} \, k_{\rm x} - 2 \eta g_{\rm eff} \, ( k_5 - 2n -1) - \nu \, k_2 - \delta_{\rm eff} \, k_3 \notag \\
&& - {1 \over 2} \kappa \, k_4 \, , \notag \\ 
\dot k_5 &=& - 2  g_{\rm eff} \, k_{\rm y} +  2 \eta g_{\rm eff} \, k_1 - 2 \delta_{\rm eff} \, k_6 - \kappa \, k_5 \, , \notag \\
\dot k_6 &=& 2 g_{\rm eff} \, k_{\rm w} +  2 \eta g_{\rm eff} \, k_2 + 2 \delta_{\rm eff} \, k_5 - \kappa \, k_6  \, , \notag \\
\dot k_7 &=& - 2 \eta g_{\rm eff} \, k_4 - 2 \nu \, k_8 \, , \notag \\
\dot k_8 &=& - 2 \eta g_{\rm eff} \, k_2 + 2 \nu \, k_7 \, ,
\end{eqnarray}
while 
\begin{eqnarray} \label{4888}
\dot m &=& \eta g_{\rm eff} \, k_4 \, .
\end{eqnarray}
Notice that these differential equations, the cooling equations, have been derived without further approximations. 

\subsection{Stationary state phonon number} \label{cooling3}

Calculating the stationary state phonon number $m^{\rm ss}$ can now be done by setting the right hand side of the above cooling equations equal to zero. Doing so, Eq.~(\ref{48}) yields for example 
\begin{eqnarray} \label{h100}
&& k_{\rm x}^{\rm ss} = 0  \, , ~~ k_{\rm y}^{\rm ss} = {4 g_{\rm eff} \kappa \nu \over \mu^3} \, , ~~
k_{\rm u}^{\rm ss} = {8 \eta g_{\rm eff}^2 \kappa \over \mu^3} \, , \notag \\
&& k_{\rm w}^{\rm ss} = {8 g_{\rm eff} (4 \eta^2 g_{\rm eff}^2 - \delta_{\rm eff} \nu ) \over \mu^3} 
\end{eqnarray}
with the cubic frequency $\mu^3$ defined as
\begin{eqnarray} \label{h200}
\mu^3 &\equiv & \nu (\kappa^2 + 4  \delta_{\rm eff}^2) - 16 \eta^2 g_{\rm eff}^2 \delta_{\rm eff} \, .
\end{eqnarray}
Moreover, we obtain the stationary state values
\begin{eqnarray} \label{h340}
n^{\rm ss} &=& {\eta^2 g_{\rm eff}^2 (\kappa^2 + 4 \nu^2 ) \over 2 \delta_{\rm eff} \mu^3} + {4 g_{\rm eff}^2 \nu^2 (\kappa^2 + 4 \delta_{\rm eff}^2 ) \over \mu^6} \notag \\
&& - {128 \eta^2 g_{\rm eff}^4 \nu \delta_{\rm eff} \over \mu^6} + {256 \eta^4 g_{\rm eff}^6 \over \mu^6}  \, , \notag \\
k_1^{\rm ss} &=& {\eta g_{\rm eff} \kappa (\kappa^2 + 4 \delta_{\rm eff}^2 ) \over 2 \delta_{\rm eff} \mu^3} - {64 \eta g_{\rm eff}^3 \kappa \nu \delta_{\rm eff} \over \mu^6} \notag \\
&& + {256 \eta^3 g_{\rm eff}^5 \kappa \over \mu^6} \, , \notag \\
k_2^{\rm ss} &=& {\eta g_{\rm eff} (\kappa^2 + 4 \delta_{\rm eff}^2 ) \over \mu^3} + {32 \eta g_{\rm eff}^3 \kappa^2 \nu \over \mu^6}  \, , \notag \\
k_3^{\rm ss} &=&{\eta g_{\rm eff} \over \delta_{\rm eff}} \, , ~~ k_4^{\rm ss} = k_8^{\rm ss} = 0  \, , \notag \\
k_5^{\rm ss} &=& - \frac{8 g_{\rm eff}^2 \nu^2 (\kappa^2 - 4 \delta_{\rm eff}^2 )}{\mu^6}  + {\eta^2 g_{\rm eff}^2 (\kappa^2 - 4 \delta_{\rm eff}^2 ) \over \delta_{\rm eff} \mu^3} \notag \\
&& - {256 \eta^2 g_{\rm eff}^4 \nu \delta_{\rm eff} \over \mu^6} + \frac{512 \eta^4 g_{\rm eff}^6}{\mu^6} \, , \notag \\
k_6^{\rm ss} &=& - {32 g_{\rm eff}^2 \kappa \nu^2 \delta_{\rm eff} \over \mu^6}  + {4 \eta^2 g_{\rm eff}^2 \kappa \over \mu^3} + {128 \eta^2 g_{\rm eff}^4 \kappa \nu \over \mu^6}  \, , \notag \\
k_7^{\rm ss} &=&  {\eta^2 g_{\rm eff}^2 (\kappa^2 + 4 \delta_{\rm eff}^2 ) \over \nu \mu^3} + {32 \eta^2 g_{\rm eff}^4 \kappa^2 \over \mu^6} \, , 
\end{eqnarray}
and 
\begin{eqnarray} \label{h3400}
m^{\rm ss} &=& {\kappa^2 + 4 \delta_{\rm eff}^2 \over 16 \nu \delta_{\rm eff}} + {\eta^2 g_{\rm eff}^2 (\kappa^2 - 8 \nu^2 + 16 \nu \delta_{\rm eff} + 4 \delta_{\rm eff}^2) \over 2 \nu \mu^3} \nonumber \\
&& + {\nu (\kappa^2 + 4 \delta_{\rm eff}^2) (\nu - 2 \delta_{\rm eff}) \over 4 \delta_{\rm eff} \mu^3} + {16 \eta^2 g_{\rm eff}^4 \kappa^2 \over \mu^6}  \, .
\end{eqnarray}
These equations can be checked easily by substituting them back into Eqs.~(\ref{48})--(\ref{4888}).

In this paper we are especially interested in the parameter regime of a tightly confined particle inside a relatively leaky optical cavity described by Eq.~(\ref{condiA}). This parameter regime is consistent with the Lamb-Dicke approximation in Eq.~(\ref{condi0}). Calculating $m^{\rm ss}$ up to second order in $\eta^2$ correctly would require to go beyond the Lamb-Dicke approximation and to take terms proportional $\eta^2$ in the system Hamiltonian into account. The above expression for the stationary state phonon number hence applies only up to first order in $\eta$. Taking this into account, Eq.~(\ref{h3400}) simplifies to 
\begin{eqnarray} \label{h345}
m^{\rm ss} &=&  {\kappa^2 + 4 (\nu- \delta_{\rm eff})^2 \over 16 \nu \delta_{\rm eff} } \, .
\end{eqnarray}
Since Eq.~(\ref{condiA}) does not pose a condition on the size of the effective detuning $\delta_{\rm eff}$, this parameter can now be used to minimise the stationary state phonon number $m^{\rm ss}$ in Eq.~(\ref{h345}). 

\begin{figure}[t]
\begin{minipage}{\columnwidth}
\hspace*{-2.1cm} \includegraphics[scale=1.35]{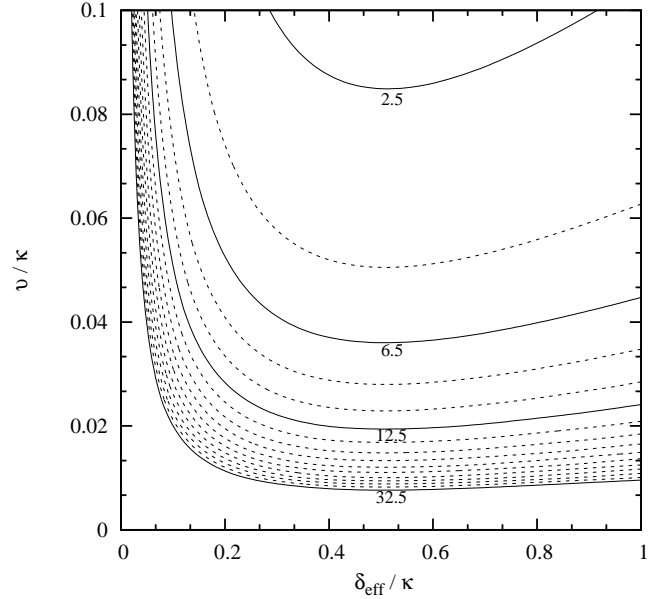}
\caption{Logarithmic contour plot of the stationary state phonon number $m^{\rm ss}$ in Eq.~(\ref{h3400}) as a function of $\delta_{\rm eff}$ and $\nu$ for $\eta = 0.1$ and $g_{\rm eff} = 0.0001 \, \kappa$. The result is in very good agreement with the simpler expression in Eq.~(\ref{h345}).} \label{geffsteadystate} 
\end{minipage}
\end{figure}

\begin{figure}[t]
\begin{minipage}{\columnwidth}
\hspace*{-2.1cm} \includegraphics[scale=1.35]{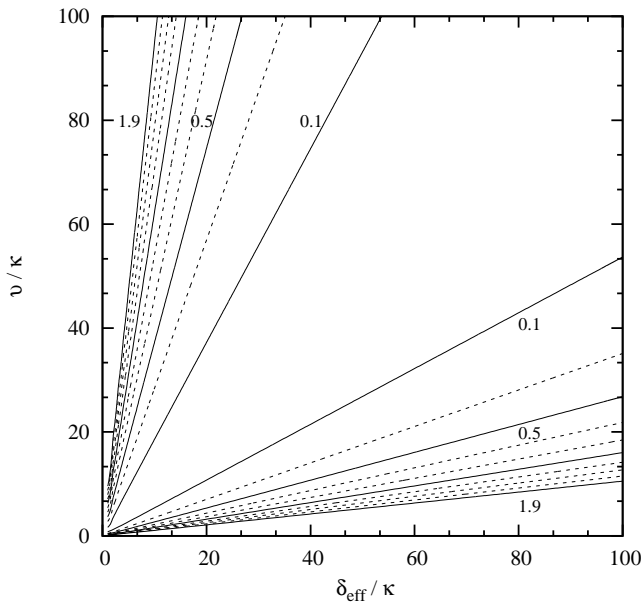}
\caption{Contour plot of the stationary state phonon number $m^{\rm ss}$ in Eq.~(\ref{h3400}) as a function of $\delta_{\rm eff}$ and $\nu$ for the same $\eta$ and $g_{\rm eff} $ as in Figure \ref{geffsteadystate} but for much larger phonon frequencies $\nu$. The result is again in very good agreement with Eq.~(\ref{h345}).} \label{geffsteadystate3} 
\end{minipage}
\end{figure}

Calculating the derivative of $m^{\rm ss}$ with respect to $\delta_{\rm eff}$ we find that the optimal choice for $\delta_{\rm eff}$ is
\begin{eqnarray} \label{deltaeff}
\delta_{\rm eff} &=& {1 \over 2} \sqrt{\kappa^2 + 4 \nu^2} \, .
\end{eqnarray}
In the parameter regime which is the most interesting from an experimental point of view, i.e.~in the case of relatively small phonon frequencies $\nu$ (weak confinement regime), the effective detuning in Eq.~(\ref{deltaeff}) becomes
\begin{eqnarray} \label{deltaeff2}
\delta_{\rm eff} = {1 \over 2} \, \kappa \, .
\end{eqnarray}
Substituting this detuning into Eq.~(\ref{h345}), we obtain the stationary state phonon number 
\begin{eqnarray} \label{mssdouble3}
m^{\rm ss}_{\delta_{\rm eff} = {1 \over 2} \, \kappa} &=& {\kappa \over 4 \nu}
\end{eqnarray}
which is in good agreement with Figure \ref{geffsteadystate}. In the other extreme case, i.e.~when the phonon  frequency $\nu$ is much larger than the cavity decay rate $\kappa$ (strong confinement regime), the effective detuning in Eq.~(\ref{deltaeff}) simplifies to
\begin{eqnarray} \label{deltaeff3}
\delta_{\rm eff} = \nu  
\end{eqnarray}
which corresponds to 
\begin{eqnarray} \label{mssdouble4}
m^{\rm ss}_{\delta_{\rm eff} = \nu} &=& {\kappa^2 \over 16 \nu^2} \, .
\end{eqnarray}
This result is confirmed by Figure \ref{geffsteadystate3} which considers much larger phonon frequencies $\nu$ than Figure \ref{geffsteadystate}. 

Comparing the two choices of effective detunings $\delta_{\rm eff}$ in Eqs.~(\ref{deltaeff2}) and (\ref{deltaeff3}), we find that cooling to very low temperatures is always limited by the relative size of the phonon frequency $\nu$ with respect to the cavity decay rate $\kappa$. However, comparing Eq.~(\ref{h345}) and Eq.~(\ref{mssdouble3}), we see that the stationary state phonon number for $\delta_{\rm eff} = {1 \over 2} \kappa$ is the square root of the stationary state phonon number for $\delta_{\rm eff} = \nu$, as it has already been pointed out in Eq.~(\ref{last}). This means, choosing the detuning $\delta_{\rm eff}$ different from the phonon frequency $\nu$, as it has been suggested in Refs.~\cite{Cirac4,vuletic3,morigi,morigi2}, yields a significant enhancement of the cavity cooling process exactly when it is most needed, namely when $m^{\rm ss}$ is larger than one. 

\subsection{Initial state}

To determine the initial state of the experimental setup in Figure \ref{setup}, we assume that the cooling laser is turned on at $t=0$. Moreover, we assume that the particle does not experience any other cooling processes and is at $t=0$ in an equilibrium state of the master equation (\ref{3.1}) but with $\Omega = 0$. Taking this into account and setting $g_{\rm eff}$ and the right hand side of the cooling equations (\ref{48})--(\ref{4888}) equal to zero, we find that this corresponds to a state with all coherences and the cavity photon number being equal to zero, i.e.
\begin{eqnarray} \label{ini}
n(0) = k_{\rm a} (0) = k_i (0) = 0 
\end{eqnarray} 
for ${\rm a} = {\rm x}, {\rm y}, {\rm u}, {\rm w}$ and $i=1, ... , 8$, while there can be any mean number of phonons $m$ in the vibrational mode of the particle. This initial condition is consistent with the particle being trapped which means that it is located around the centre of a trap and that it has no initial momentum away from its equilibrium position. The first of these two statements implies $k_{\rm u} (0) = 0$ and the second one implies $k_{\rm x} (0) = 0$.

\subsection{Cooling dynamics} \label{appp}

To calculate the effective cooling rate $\gamma$, we notice that only one of the variables in the above cooling equations, namely the mean phonon number $m$, evolves on a relatively slow time scale. All other variables, i.e. the mean photon number $n$ and the coherences, evolve on the fast time scale given by $\nu$ and $\kappa$. In the parameter regime of Eq.~(\ref{condiA}), these can hence be eliminated adiabatically from the time evolution of the system, leaving us only with a single effective cooling equation. Doing so and setting the time derivatives in Eqs.~(\ref{48}) and (\ref{49}) equal to zero and assuming that we are at the beginning of the cooling process where $m \gg 1$, we find 
\begin{eqnarray} \label{k444}
k_4 &=& - {64 \eta g_{\rm eff} \nu \kappa \delta_{\rm eff} \over (\kappa^2 + 4 \nu^2)^2 + 8 \delta_{\rm eff}^2 (\kappa^2 - 4 \nu^2) + 16 \delta_{\rm eff}^4} \, m  \, . ~~~~
\end{eqnarray}
This equation holds up first order in $\eta g_{\rm eff}$ and is consistent with the Lamb-Dicke approximation introduced in Section \ref{simple}. 

\begin{figure}[t]
\begin{minipage}{\columnwidth}
\hspace*{-1cm} \includegraphics[scale=0.7]{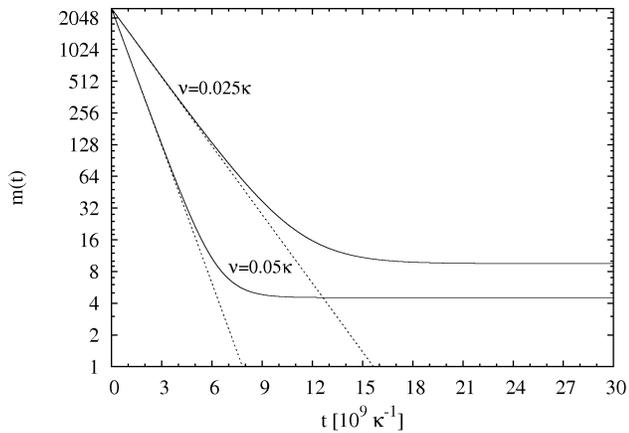}
\caption{Logarithmic plot of the time evolution of the mean phonon number $m$ for $\eta = 0.1$, $g_{\rm eff} = 0.0005 \, \kappa$, and $\delta_{\rm eff} = {1 \over 2} \, \kappa$ for two different phonon frequencies $\nu$. The solid lines have been obtained from a numerical solution of the cooling equations (\ref{48})--(\ref{4888}) for the initial conditions in Eq.~(\ref{ini}) and $m(0) = 2500$. The dashed lines assume an exponential cooling process with the rate $\gamma $ in Eq.~(\ref{gammasingle2}). Both solutions coincide very well when $m$ is far away from its stationary state value.} \label{kuCyanBlue2} 
\end{minipage}
\end{figure}

\begin{figure}[t]
\begin{minipage}{\columnwidth}
\hspace*{-1cm} \includegraphics[scale=0.7]{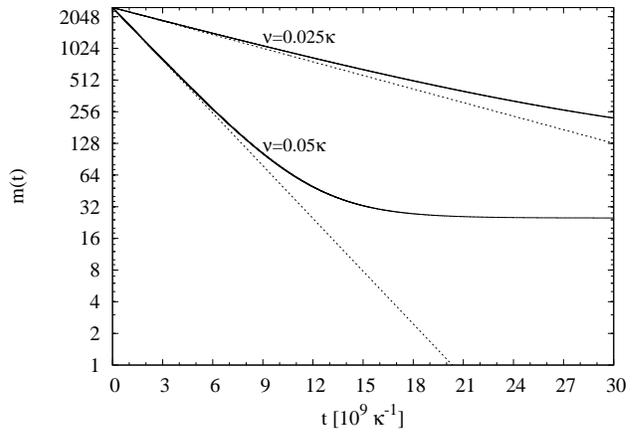}
\caption{Logarithmic plot of the time evolution of the mean phonon number $m$ for the same experimental parameters as in Figure \ref{kuCyanBlue2} but with $\delta_{\rm eff} = \nu$. Again, the solid lines have been obtained from a numerical solution of the cooling equations (\ref{48})--(\ref{4888}) for the initial conditions in Eq.~(\ref{ini}) and $m(0) = 2500$. The dashed lines assume an exponential cooling process with the cooling rate $\gamma$ in Eq.~(\ref{gammasingle2}). Although $\kappa$ and $\nu$ remain the same, we now observe slower cooling processes with higher stationary state phonon numbers $m^{\rm ss}$.} \label{kuCyanBlue} 
\end{minipage}
\end{figure}

Eq.~(\ref{4888}) shows that the photon-phonon coherence $k_4$ is essentially the cooling rate of the trapped particle. Substituting Eq.~(\ref{k444}) into Eq.~(\ref{4888}), we obtain the final cooling equation
\begin{eqnarray} \label{rates2}
\dot m &=& - \gamma \, m 
\end{eqnarray}    
with the cooling rate $\gamma$ given by
\begin{eqnarray} \label{gammasingle2}
\gamma &=& {64 \eta^2 g_{\rm eff}^2 \nu \delta_{\rm eff} \kappa \over (\kappa^2 + 4 \nu^2)^2 + 8 \delta_{\rm eff}^2 (\kappa^2 - 4 \nu^2) + 16 \delta_{\rm eff}^4} \, . ~~~~
\end{eqnarray}
Solving this equation yields
\begin{eqnarray} \label{mexact}
m(t) &=& {\rm e}^{- \gamma t} \, m(0) \, .
\end{eqnarray}    
Since the rate $\gamma$ is always positive, this equations describes an exponential reduction of the initial phonon number $m (0)$, i.e.~cooling. Figures \ref{kuCyanBlue2} and \ref{kuCyanBlue} compare the exponential cooling process with the rate $\gamma$ in Eq.~(\ref{gammasingle2}) (dashed lines) with an exact numerical solution of the full set of cooling equations (solid lines) and shows that  $\gamma$ is indeed a very good approximation for the cooling rate as long as the actual phonon number $m$ is much larger than one. The exponential reduction of $m$ only slows down when $m$ approaches its stationary state value. We also see that the speed of the cooling process increases, as $\nu / \kappa$ increases. 

As in the previous section, let us now have a closer look at the two extreme cases, where the phonon frequency $\nu$ is either much smaller or much larger than the cavity decay rate $\kappa$. In the first case (weak confinement regime), one should choose $\delta_{\rm eff} = {1 \over 2} \kappa$ (c.f.~Eq.~(\ref{deltaeff2})) in order to minimise the stationary state phonon number. Doing so, Eq.~(\ref{gammasingle2}) simplifies to 
\begin{eqnarray} \label{gammasingle3}
\gamma_{\delta_{\rm eff} = {1 \over 2} \kappa} &=& {8 \eta^2 g_{\rm eff}^2 \nu \kappa^2 \over \kappa^4 + 4 \nu^4} \, . ~~~~
\end{eqnarray}
In the latter case (strong confinement regime), one should choose $\delta_{\rm eff} = \nu$ (c.f.~Eq.~(\ref{deltaeff3})) in order to minimise the stationary state phonon number. This choice of $\delta_{\rm eff}$ implies 
\begin{eqnarray} \label{gammasingle4}
\gamma_{\delta_{\rm eff} = \nu} &=& {64 \eta^2 g_{\rm eff}^2 \nu^2 \over \kappa (\kappa^2 + 16 \nu^2)} \, . ~~~~
\end{eqnarray}
Both cooling rates scale as $\eta^2 g_{\rm eff}^2$ and are hence in general relatively small (c.f.~Eq.~(\ref{condiA})). As already pointed out in Eq.~(\ref{gammadouble2}), comparing the analytical expressions in Eqs.~(\ref{gammasingle3}) and (\ref{gammasingle4}) we find that $\gamma_{\delta_{\rm eff} = {1 \over 2} \kappa}$ is about $\kappa / 8 \nu$ times larger than $\gamma_{\delta_{\rm eff} = \nu}$. This is confirmed by Figures \ref{kuCyanBlue2} and \ref{kuCyanBlue} which show that choosing $\delta_{\rm eff} = {1 \over 2} \kappa$ in the weak confinement regime not only leads to a lower stationary state phonon number but also to a significant speed up of the cooling process. Large cooling rates are important when the purpose of using a cavity is to avoid spontaneous emission from the particle.

\subsection{Avoiding spontaneous emission from the particle}

The analysis of the cooling process presented in this paper only applies, when the population in the excited atomic state $|1 \rangle$ remains negligible. Otherwise, the adiabatic elimination in Section \ref{adel} does not hold and the cooling process becomes a mixture of cavity and ordinary laser cooling \cite{vuletic3}. Avoiding spontaneous emission from the excited electronic state  $|1 \rangle$ is especially important when it comes to the cooling of molecules, where it could result in the population of states, where the particle no longer experiences the cooling laser. We therefore conclude this section with an estimation of the parameter regime, where spontaneous emission from the particle remains highly unlikely. 

In the Lamb-Dicke limit and the parameter regime given by Eq.~(\ref{condi}), Eq.~(\ref{c1-c0}) shows that the population in $|1 \rangle$ scales essentially as $\Omega^2/\Delta^2$. We therefore assume in the following that 
\begin{eqnarray} \label{condi3}
\gamma &\gg & {\Gamma \Omega^2 \over 4 \Delta^2} \, ,
\end{eqnarray}
i.e.~that the cooling rate is much larger than the probability density for the spontaneous emission of a photon from the particle. For $\delta_{\rm eff} = {1 \over 2} \kappa$ and $\gamma$ as in Eq.~(\ref{gammasingle3}) and when taking the definition of $g_{\rm eff}$ in Eq.~(\ref{3.8}) into account, we see that this condition applies when
\begin{eqnarray} \label{condi4}
{g^2 \over \kappa \Gamma} &\gg & {\kappa^4 + 4 \nu^4 \over 8 \eta^2 \nu \kappa^3} \, .
\end{eqnarray}
Since $\eta \ll 1$, the right hand side of this equation is in general much larger than one. This means, spontaneous emission from the particle is only negligible, when the cavity is operated in the so-called strong coupling regime. If the cavity decay rate $\kappa$ is much smaller than $4 \nu$, one should choose $\delta_{\rm eff} = \nu$ and the cooling rate simplifies to the expression in Eq.~(\ref{gammasingle4}). In this case, condition (\ref{condi3}) simplifies to
\begin{eqnarray} \label{condi5}
{g^2 \over \kappa \Gamma} &\gg & {\kappa^2 + 16 \nu^2 \over 64 \eta^2 \nu^2} \, .
\end{eqnarray}
Since $\eta \ll 1$, we find again that the cavity needs to be operated within the strong coupling regime. 

\section{Comparison with ordinary laser cooling} \label{new}

In the parameter regime of a tightly confined particle inside a relatively leaky optical cavity described by Eq.~(\ref{condiA}), the cavity cooling scenario in Figure \ref{setup} has many similarities with ordinary laser cooling \cite{sideband,Stenholm2,sideband2,ions}. The reason is that the atomic 0--1 transition and the cavity are so strongly detuned that the electronic states of the trapped particle can be adiabatically eliminated from the system dynamics (c.f.~Section \ref{adel}). The remaining master equation (c.f.~Eq.~(\ref{3.1})) with the interaction Hamiltonian $H_{\rm I}$ in Eq.~(\ref{3.7b}) is almost the same as in laser cooling. One only needs to replace the cavity annihilation operator $c$ by the atomic lowering operator $|0 \rangle \langle 1|$, the cavity decay rate $\kappa$ by the atomic decay rate $\Gamma$, and the effective coupling constant $g_{\rm eff}$ by the cooling laser Rabi frequency $\Omega$, and so on. A crucial advantage of cavity cooling compared to ordinary laser cooling could be the lack of recoil heating. However, this difference does not matter as long as there is always only a very small number of photons inside the cavity.

\subsection{Comparison of cavity cooling with ordinary laser cooling}

As in laser cooling, we found that it is useful to distinguish between two different parameter regimes: the strong confinement regime and the weak confinement regime. In the strong confinement regime, where the relevant spontaneous decay rate ($\kappa$ or $\Gamma$) is much smaller than the phonon frequency $\nu$, one should choose the relevant laser detuning equal to $\nu$ (c.f.~Eq.~(\ref{deltaeff3})) in order to minimise the stationary state phonon number $m^{\rm ss}$. In laser cooling, this case is known as {\em laser sideband cooling}. The stationary state phonon number $m^{\rm ss}$ for cavity cooling in the strong confinement regime is essentially given by $\kappa^2/16 \nu^2$ (c.f.~Eq.~(\ref{h345})) while scaling as $\Gamma^2/ \nu^2$ in laser sideband cooling \cite{ions}. This means, it is possible to cool to phonon numbers well below one (ground state cooling), although realising $\kappa \ll \nu$ in cavity cooling is experimentally very demanding. 

In the weak confinement regime, where the relevant spontaneous decay rate ($\kappa$ or $\Gamma$) is much larger than the phonon frequency $\nu$, one should choose the relevant laser detuning equal to ${1 \over 2} \kappa$ (c.f.~Eq.~(\ref{deltaeff2})) or ${1 \over 2} \Gamma$, respectively, in order to minimise the stationary state phonon number $m^{\rm ss}$.
In this case, the stationary state phonon number $m^{\rm ss}$ equals $\kappa/4 \nu$ (c.f.~Eq.~(\ref{mssdouble3})) in cavity cooling while it scales as $\Gamma/\nu$ in laser cooling \cite{ions}. This is exactly what one would expect when the described cavity cooling scheme and ordinary laser cooling are more or less equivalent.

\subsection{Comparing $\delta_{\rm eff} = \nu$ with $\delta_{\rm eff} = {1 \over 2} \kappa$ in cavity cooling} \label{comp}

\begin{figure}[t]
\begin{minipage}{\columnwidth}
\hspace*{-2.1cm} \includegraphics[scale=1.35]{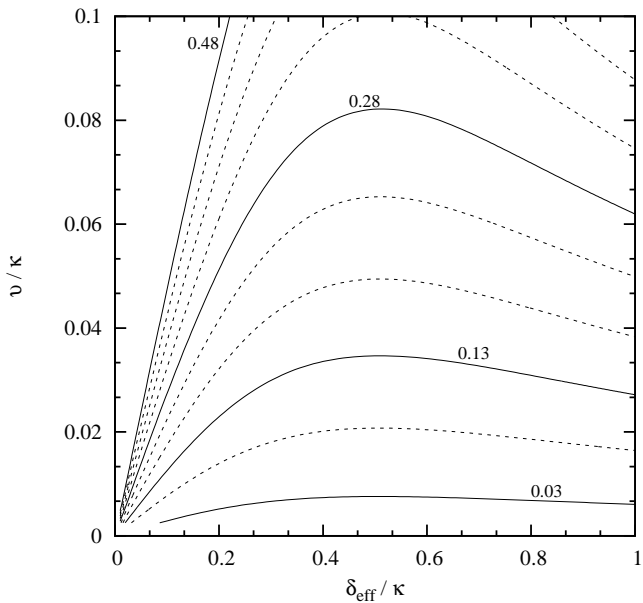}
\caption{Contour plot of $m^{\rm ss}/m^{\rm ss}_{\delta_{\rm eff} = \nu} $ as a function of $\delta_{\rm eff}$ and $\nu$. This plot has been obtained using Eq.~(\ref{h3400}) for $\eta = 0.1$ and $g_{\rm eff} = 0.0001 \, \kappa$ and shows that choosing the detuning $\delta_{\rm eff}$ comparable to $\kappa$ leads to much lower stationary state phonon numbers than when $\nu \ll \kappa$.} \label{geffsteadystate2} 
\end{minipage}
\end{figure}

\begin{figure}[t]
\begin{minipage}{\columnwidth}
\hspace*{-2.1cm} \includegraphics[scale=1.35]{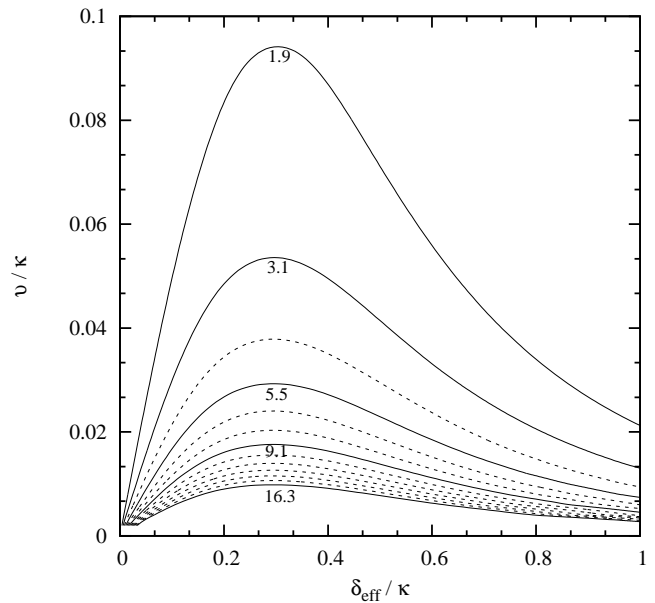}
\caption{Contour plot of $\gamma/\gamma_{\delta_{\rm eff} = \nu} $ as a function of $\delta_{\rm eff}$ and $\nu$ for the same parameters as in Figure \ref{geffsteadystate2}. This plot has been obtained using Eq.~(\ref{gammasingle2}) and shows that  
that choosing the detuning $\delta_{\rm eff}$ comparable to $\kappa$ leads to a significant speed up the cooling process when $\nu \ll \kappa$.} \label{c4_2D} 
\end{minipage}
\end{figure}

Previous papers (see e.g.~Refs.~\cite{Cirac4,morigi,morigi2,vuletic3}) mainly focus their analysis on cavity cooling in the strong confinement regime, where one should choose $\delta_{\rm eff} = \nu$ and where it is in principle possible to cool the trapped particle to phonon numbers well below one. The purpose of this paper is to point out that there are three distinct advantages in choosing $\delta_{\rm eff}$ differently, i.e.~close to ${1 \over 2} \kappa$ (c.f.~Eq.~(\ref{deltaeff})), when it is experimentally not possible to enter the strong confinement regime:
\begin{enumerate}
\item A reduction of the stationary state phonon number. As already pointed out in Eq.~(\ref{last}), $m^{\rm ss}_{\delta_{\rm eff} = {1 \over 2} \kappa}$ equals the square root of the stationary state phonon number $m^{\rm ss}_{\delta_{\rm eff} = \nu}$ and is hence significantly smaller than $m^{\rm ss}_{\delta_{\rm eff} = \nu}$ for a wide range of experimental parameters. This result is confirmed by Figure \ref{geffsteadystate2} which shows $m^{\rm ss} / m^{\rm ss}_{\delta_{\rm eff} = \nu}$ for relatively small phonon frequencies $\nu$ and a wide range of effective detunings $\delta_{\rm eff}$. In order to minimise the stationary state phonon number, one should choose $\delta_{\rm eff}$ as in Eq.~(\ref{deltaeff}).
\item An increase of the cooling rate. Calculating the ratio $\gamma_{\delta_{\rm eff} = {1 \over 2} \kappa}/\gamma_{\delta_{\rm eff} = \nu} $ using Eqs.~(\ref{gammasingle3}) and (\ref{gammasingle3}) for fixed values of $\eta$, $g_{\rm eff}$, $\kappa$, and $\nu$, we find 
\begin{eqnarray}
{\gamma_{\delta_{\rm eff} = {1 \over 2} \kappa} \over \gamma_{\delta_{\rm eff} = \nu}} &=& {\kappa \over 2 \nu} \cdot {\kappa^2 (\kappa^2 + 16 \nu^2) \over 4 \kappa^4 + 16 \nu^4}
\end{eqnarray}
which scales approximately as $\kappa / 8 \nu$, as already pointed out in Eq.~(\ref{gammadouble2}). This means, choosing $\delta_{\rm eff}$ close to ${1 \over 2} \kappa$ results in a significant speedup of the cooling process. This result is confirmed by Figure \ref{c4_2D} which shows $\gamma / \gamma_{\delta_{\rm eff} = \nu}$ for the same parameters as in Figure \ref{geffsteadystate2}.
\item Minimising spontaneous emission from the excited electronic state $|1 \rangle$ of the trapped particle. This is important, when it comes for example to the cooling of molecules, where such an emission might populate states, where the particle no longer experiences the cooling laser. As pointed out in the previous paragraph, the cooling rate $\gamma$ is much higher when $\delta_{\rm eff}$ is close to ${1 \over 2} \kappa$. As a consequence, the 
restrictions which need to be posed on the minimum size of the single particle cooperativity parameter $g^2 / \kappa \Gamma$ are therefore much weaker in this case. The reduction of the cooling time might moreover help to balance unconsidered heating processes which are, for example, due to stray fields.
\end{enumerate}

\section{Conclusions} \label{conc}

This paper analyses cavity cooling of a trapped particle with ground state $|0 \rangle$ and excited state $|1 \rangle$ in the parameter regime, where the phonon frequency $\nu$ and the cavity decay rate $\kappa$ are both relatively large. As shown in Figure \ref{setup}, we assume that the motion of the particle orthogonal to the cavity axis, i.e.~in the direction of the cooling laser, is strongly confined. The aim of the cooling process is to minimise the number of phonons in this vibrational mode. The role of the cooling laser is to establish a coupling between the phonons and the electronic states of the trapped particle. The result is the continuous conversion of phonons into cavity photons which subsequently leak into the environment. The finite temperature limit of the cooling process is due to the counter-rotating terms in the system Hamiltonian which simultaneously create a phonon and a cavity photon.

We describe the time evolution of the system using the usual master equation. The main approximations made in the derivation of the master equation are the Lamb-Dicke approximation, the rotating wave approximation with respect to optical transitions, and the adiabatic elimination of the excited atomic state $|1 \rangle$ of the particle. The second approximation is a standard approximation in quantum optics and in general in very good agreement with experimental findings. The third approximation requires a sufficiently large detuning $\Delta$ of the 0--1 transition of the particle with respect to the cavity field and the cooling laser (c.f.~Figure \ref{energy1}). Without further approximations, the master equation can then be used to obtain a closed set of differential equations, the cooling equations, which predict the time evolution of the mean phonon number. To calculate the stationary state phonon number, we set the right hand side of the cooling equations equal to zero. The cooling rate is obtained via an approximate solution of the cooling dynamics with the help of an adiabatic elimination which applies for relatively large $\kappa$, $\nu$, and $m $ and is in very good agreement with numerical simulations.

Our results show that, in the case of a relatively tightly confined particle inside a relatively leaky optical cavity, the cavity cooling scenario in Figure \ref{setup} is essentially equivalent to ordinary laser cooling. This applies, since the mean number of photons inside the cavity remains negligible in the parameter regime considered here. If the purpose of the cavity is to avoid photon emission from the excited electronic state $|1 \rangle$ of the trapped particle, as it might be necessary for the cooling of molecules, the cavity needs to be operated well within the strong coupling regime. Otherwise, the cooling process described here simply becomes a mixture of cavity and ordinary laser cooling. 

Before concluding this paper we would like to point out that the above results which have also been obtained by other authors are only valid within the Lamb-Dicke regime. It is usually believed that the Lamb-Dicke approximation, i.e.~the expansion of the system Hamiltonian as a function of $\eta$, is well justified for a sufficiently pre-cooled particle. However, when avoiding this approximation, the methods used in this paper no longer yield a closed set of rate equations. In fact, higher-order terms in $\eta$ which have been neglected here could, in principle, evolve the system in the long term into a state which is very different from the stationary state obtained above \cite{tony2}. When calculating stationary states, a priori assumptions about the existence and the $\eta$-dependence of the stationary state phonon number should be avoided. Moreover, going beyond the Lamb-Dicke regime yields corrections to the cooling rate $\gamma$ of the order $\eta^2$ \cite{morigi}. \\[0.5cm]

{\em Acknowledgement.} The authors thank Wolfgang Alt for pointing out a serious problem in the first version of this paper. A. B. acknowledges a James Ellis University Research Fellowship from the Royal Society and the GCHQ. This work was supported by the UK Research Council EPSRC and the European Union ESF EuroQUAM programme CMMC.

\end{document}